\documentclass[Afour,sageh,times]{sagej}

\usepackage{moreverb,url,array}

\usepackage[colorlinks,bookmarksopen,bookmarksnumbered,citecolor=red,urlcolor=red]{hyperref}

\newcommand\BibTeX{{\rmfamily B\kern-.05em \textsc{i\kern-.025em b}\kern-.08em
T\kern-.1667em\lower.7ex\hbox{E}\kern-.125emX}}
\setcitestyle{numbers}

\def\volumeyear{2024}

\begin{document}

\runninghead{Rabb, Ozkan}

\title{Beyond case studies: Teaching data science critique and ethics through sociotechnical surveillance studies}

\author{Nicholas Rabb\affilnum{1} and Desen Ozkan\affilnum{2}}

\affiliation{\affilnum{1}Department of Computer Science, Tufts University, US\\
\affilnum{2}Center for Engineering Education and Outreach, Institute for Research on Learning and Instruction, Tufts University, US}

\corrauth{Nicholas Rabb
Tufts University,
Medford, MA 02145,
US.}

\email{nicholas.rabb@tufts.edu}

\begin{abstract}

Ethics have become an urgent concern for data science research, practice, and instruction in the wake of growing critique of algorithms and systems showing that they reinforce structural oppression. There has been increasing desire on the part of data science educators to craft curricula that speak to these critiques, yet much ethics education remains individualized, focused on specific cases, or too abstract and unapplicable. We synthesized some of the most popular critical data science works and designed a data science ethics course that spoke to the social phenomena at the root of critical data studies -- theories of oppression, social systems, power, history, and change -- through analysis of a pressing sociotechnical system: surveillance systems. Through analysis of student reflections and final projects, we determined that at the conclusion of the semester, all students had developed critical analysis skills that allowed them to investigate surveillance systems of their own and identify their benefits, harms, main proponents, those who resist them, and their interplay with social systems, all while considering dimensions of race, class, gender, and more. We argue that this type of instruction -- directly teaching data science ethics alongside social theory -- is a crucial next step for the field.

\end{abstract}


\maketitle

\section{Introduction}

Ethics have become a popular and urgent consideration for data science research, practice, and instruction. In the wake of several high-profile critiques of algorithms and systems -- including showing that they reinforce structural inequalities \cite{ONeil2016WeaponsDemocracy}, reproduce racism \cite{Noble2018AlgorithmsRacism} or work in a racist manner \cite{Buolamwini2018Gender,Benjamin2019RaceCode}, and inevitably encode other oppressive biases \cite{DIgnazio2020DataFeminism} -- there has been increasing desire on the part of data science educators to craft curricula that speak to these critiques. As such, data science and engineering programs across the U.S. have begun teaching ethics through various means, including teaching philosophical theories of ethical decision-making, and investigating ethical case studies that often encode dilemmas \cite{Hess2019ConditionsCases}. Yet review of ethical instruction reveals that there are gaps, namely that the ethics are too individualized, and that deep understanding of the knowledge that enabled these influential critiques in the first place is missing.

Through crafting a course for the Tufts University computer science department, we developed a conception of ethical instruction that addresses these areas. By synthesizing several of the major critical data science works, we arrived at the conclusion that social knowledge is what is missing in ethical instruction: understandings of oppression, social systems and their logics, power, history, and theories of change. Knowledge of these aspects of critical theory, or lived experience of their effects, have enabled data science critics to describe a higher level of ethical issue with data science -- one that is generative and able to detect many issues with the diverse manifestations of data technology. The popular data science ethics that are taught -- fairness, accountability, transparency, philosophical theories, ethical dilemma case studies -- are only encoding ethical instruction in so far as they are speaking to these larger systems. Thus, they stand as crucial to include in data science education.

Our course analyzed surveillance systems along the lines of social knowledge we identified, and attempted to impart ethical skills to students through practicing their ability to critique and reimagine sociotechnical systems. Through sections involving deconstructing what the problems of surveillance systems are, learning the practice of critique that led to identifying those problems, and applying those skills to a surveillance technology of their own choosing, the course exercised students' muscles of critique. Additionally, content was grounded in contemporary examples of surveillance systems, discussing surveillance capitalism and the harvesting of personal data, technological surveillance in the workplace, data collection systems that assist in border policing, as well as systems that aid local policing. At the semester's end, students were able to critique a surveillance system of their own choosing, demonstrating their ability to analyze sociotechnical systems from the standpoint of oppression, social logics, power, history, and change.

To assess the degree of success of our course, we collected student reflection assignments and course feedback through a consent process involving the Tufts IRB, and qualitatively analyzed their responses. We found that students were enthusiastic about what they had learned, felt that their concepts of technology and ethics had been challenged by the material, and that they found the content highly relevant to their own lives. Their reflections and final projects demonstrated that each student had developed a facility with critical analysis that reflected a deep level of nuanced ethical understanding. Reflecting on our success, we recommend that this type of instruction be made an integral part of the ethical education of data science students, and also that research beyond our preliminary qualitative analysis be done in other ethical learning courses.

\section{The contextual nature of ethics}

Ethical analysis of data science algorithms has become an urgent topic of study in the wake of several high-profile critiques of widely used data-driven systems, demonstrating that their mechanisms reproduce harms across lines of, for example, racial \cite{Buolamwini2018Gender,Benjamin2019RaceCode}, gender \cite{DIgnazio2020DataFeminism}, and economic oppression \cite{ONeil2016WeaponsDemocracy}. In response, the data science community has sought methods to make algorithms more ethical. The dominant trend in the data science community's embrace of ethics is to de-bias algorithms; that is, consider their \emph{fairness}, \emph{accountability}, and \emph{transparency} \cite{DIgnazio2020DataFeminism,Costanza-Chock2020DesignNeed}. The prevailing attitude is that ethical algorithms should treat all users equally, include mechanisms where their logic can be rewritten if needed, and make public that logic. This is a significant step forward for data science, as no such ethical framework was widely recognized not long ago.

This trend has motivated university educators to include these concepts in turn, and has even been incorporated into the Accreditation Board for Engineering and Technology (ABET), which vets university ethics programs. Hess and Fore, through a review of engineering ethics interventions, determined that, to teach engineers how to craft fair, accountable, or transparent systems, ethics education primarily focuses on individual decision-making as ethics \cite{Hess2019ConditionsCases}. Particularly, many programs teach ethical skills using ethical heuristics -- often through case studies -- and philosophical ethics.



Tufts University has followed suit, with several ethics courses having been offered in recent years. The Department of Computer Science has featured \emph{Ethical Issues in Computer Science and Technology}, which introduced discussions of case studies presenting ethical dilemmas generated by biased algorithms. Students learn ethical principles from philosophy, and have discussions centered around identifying biases and arguing for mechanisms to promote fairness, accountability, and transparency. Another course, \emph{Ethics for AI, Robotics, and Human Robot Interaction}, encourages students to consider ethical dilemmas through the lenses of classical philosophical theories, including consequentialism, deontology, and virtue ethics. In the Fall 2021 semester, a course co-listed in Computer Science and the Fletcher School of Diplomacy, titled \emph{AI, Ethics, and Policy}, similarly encouraged discussion of biased algorithms responsible for making social decisions, and policies that seek to remedy these harms. These courses represent a significant step forward for the university and its ethical instruction of computing and data science.

Yet the case studies presented in the classroom and the ethical theories that guide students to solutions are missing crucial histories and sociopolitical context. For one, abstract ethical theories become dependent on social understanding when applied to real situations. For example, consequentialism, which is concerned with assessing ethics in terms of harms and benefits, is dependent on a knowledge of what harms or benefits even exist in a given scenario. Fairness depends on a definition of ``fair,'' which is by no means obvious, and subsequently relies on concepts of justice. Likewise for deontological edicts, the virtues guiding virtue ethics, or considerations of accountability and transparency, there is inevitably a social worldview that drives reasoning towards ethical goals. While case studies present ethical dilemmas involving harm or bias, the interpretation of the harms is entirely subject to the worldview of the interpreter.

To illustrate this, consider several examples of different views of fairness, harm, virtue, and consequence, and the ethical interpretations they lead to. If students are considering an algorithm intended to judge criminal recidivism (the risk that an incarcerated person will be incarcerated again), their ability to identify potential harm rests, for example, on their understanding of systemic racism \cite{Benjamin2019RaceCode}. A worldview steeped in white supremacy or anti-Black racism may not consider the algorithm harmful, but rather a just treatment of criminals, which would then inform any ethical assessments. For another example, an algorithm guiding a lethal autonomous weapon will have different ethical dilemmas identified based on a student's ideology surrounding American Exceptionalism, their understanding of foreign policy, or their views on imperialism. This is not just true for students, but also for instructors, whose selection of case studies and assessment of them is influenced by their worldview.

\begin{figure}[h]
    \centering
    \includegraphics[width=\linewidth]{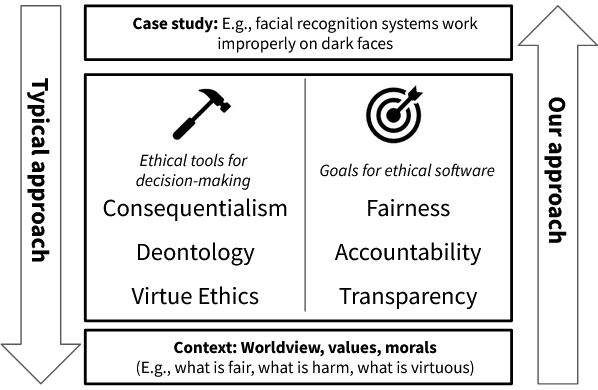}
    \caption{An illustration of the typical approach to data science ethics (top-down from case studies) versus our approach (bottom up from worldview and context).}
    \label{fig:ethical-approaches}
\end{figure}

Taking this into consideration, it is therefore no surprise that many of the critiques of data science and case studies that have been taught in classrooms are the product of those scholars who are well-versed in understanding the worldview itself -- those who are employing critical social theory either through training or lived experience. To list a few examples, Ruha Benjamin, whose book \emph{Race After Technology: Abolitionist Tools for the New Jim Code} \cite{Benjamin2019RaceCode} has sparked discussion of anti-Black racism in technology, completed a PhD in sociology and is a professor of African American Studies. Joy Buolamwini, who critiqued facial recognition technology's inability to recognize Black faces, is steeped in the history of anti-Black racism, as her popular work ``AI, Ain't I A Woman?'' \cite{Buolamwini2018AIWoman} is a reference to Sojourner Truth's famous 1851 speech at the Women's Convention ``Ain't I A Woman?''. Cathy O'Neil, whose work titled \emph{Weapons of Math Destruction} illustrates several case studies of unjust algorithms, was involved in social movement organizing with Occupy Wall Street \cite{Smith2012TheONeil}. These scholars have brought data ethics into the spotlight because their understandings of the social world enabled them to generate critiques and identify harms.

Thus, there is an argument to expose data science students to the same theoretical and social knowledge that allows critical thinkers to identify ethical issues. This instruction complements current instruction on higher-level ethical considerations. The context surrounding discussions of ethical philosophy and case studies deserves to be made explicit, which requires instruction that dips into the social studies.

\section{Grounding ethics in social study}

There is a level of ethical study that operates at the root of conceptual understanding, probing at and teaching ideology and different worldviews. This type of study involves learning about major social systems such as capitalism, socialism, neoliberalism, democracy; theories of oppression along racial, class, or gender lines; historical analysis and social change theory; and much more. Learning the vocabulary and concepts necessary to understand these abstract (yet also very concrete) bodies of work enables ethical judgment at a general level. Instruction through case studies, in contrast, may only trigger ethical alarms with very similar situations. The process of generalization is key to learning and identifying new instances of a pattern, so learning critical social studies is crucial for allowing students to reason ethically across many domains and cases.

Currently, we identified several popular critical branches of thought that are informing widespread concern for ethical data science. Each is an example of \emph{critical theory}, that which aims to interrogate the current state of the world and its systems, asking if they are justified or not, and investigating their effects towards the oppression or liberation of all people \cite{James2021CriticalTheory}. Among them are theories of power, theories of oppression, theories of social systems and their logics, historical analysis, and theories of change.

\begin{figure}[h]
    \centering
    \includegraphics[width=\linewidth]{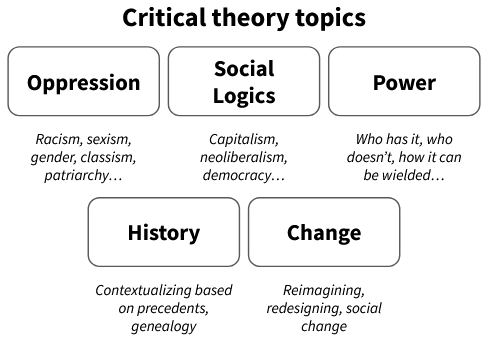}
    \caption{An illustration of the five sub-topics of critical theory we identified in major critical data science works, and aimed to teach to students.}
    \label{fig:critical-theory-topics}
\end{figure}

The most visible line of study that is in the spotlight of ethical concern with data science is the study of \emph{oppression}. Data science has become concerned with racism, gender oppression, and economic oppression. These concerns reflect major works in the area, such as Benjamin's \emph{Race After Technology} \cite{Benjamin2019RaceCode}, Buolamwini's work on facial recognition \cite{Buolamwini2018Gender}, O'Neil's \emph{Weapons of Math Destruction} \cite{ONeil2016WeaponsDemocracy}, Eubanks' \emph{Automating Inequality} \cite{Eubanks2017AutomatingPoor}, and more. These analyses of specific aspects of algorithmic injustice are crucial to bring to data science study, as they elucidate dominant logics of oppression that are often invisibly woven into society, and discuss the ways that data technologies are shaped by and reproduce them.

Other popular concerns in data science have stemmed from analysis of data science intertwined with the likes of capitalism, neoliberalism, socialism, democracy, autocracy, and other major political and economic \emph{logics}. These concerns have been brought to the field by major works such as Zuboff's \emph{Surveillance Capitalism} \cite{Zuboff2019ThePower}, Sunstein's \emph{Republic.com} \cite{Sunstein2001Republic.com}, Noble's \emph{Algorithms of Oppression} \cite{Noble2018AlgorithmsRacism}, D'Ignazio and Klein's \emph{Data Feminism} \cite{DIgnazio2020DataFeminism}, and others. It is crucial that data science practitioners understand the ethical dimensions of the economic system they work within, the political system that governs or does not govern their work, and the values associated with abstract philosophies like democracy or freedom.

Behind all of these subjects are considerations of \emph{power}: who has it, who does not, how it is wielded, and how it interfaces with data technologies. Power, despite its pervasive nature -- weaving through every consideration of bias, fairness, accountability, oppression, and social force -- is an infrequent lens through which data science is analyzed, even though it motivates the above critiques. Some notable works, including D'Ignazio and Klein's \emph{Data Feminism} \cite{DIgnazio2020DataFeminism} and Costanza-Chock's \emph{Design Justice} \cite{Costanza-Chock2020DesignNeed}, do discuss this matter explicitly, while others, such as the works mentioned above, do so more implicitly.

Another lens through which data science has been scrutinized is that of \emph{history}. Following in the footsteps of the humanities' shift to contextualizing social systems in terms of their genealogies \cite{Foucault1969TheKnowledge}, several influential critics have contextualized data science with historical analysis -- including Zuboff's situating data surveillance in the history of Western industrial capitalism \cite{Zuboff2019ThePower}, Browne's investigation of technological surveillance in conversation with U.S. slavery \cite{Browne2015DarkBlackness}, and Benjamin's comparison of racist technology with Jim Crow era practices \cite{Benjamin2019RaceCode}.
Analysis of technological history shows that each of the dimensions of ethical study listed above -- oppression, social logics, and power -- have roots, have constantly been in flux, have been shaped by active and intentional forces, and can continue to be shaped. This combats a typical view of technological determinism \cite{Zuboff2019ThePower} that imagines data technologies as being inevitable rather than manifesting because of intentional action, subject to social forces. Historical analysis has also challenged our view of what even counts as technology, showing that past non-digital technologies were subject to similar ethical problems as today as a result of their social interactions rather than the specific manifestation of mechanism.

Finally, several major critiques have embraced discussing \emph{theories of change} in the face of undesirable technologies subject to the aforementioned social forces. When ethical issues with technology move to the realm of the abstract, finding themselves defined by forces like racism, sexism, or state violence, it is no longer sufficient to characterize ethical data science with only individual actions like checking for bias or pushing for algorithms to be transparent. There are many traditions of social and political action that lend themselves to taking ethical action at the systemic or cultural level; including acts of resistance \cite{Browne2015DarkBlackness}, organizing labor \cite{McAlevey2016NoAge}, collective demonstration \cite{Engler2016ThisCentury}, legislative regulation \cite{Zuboff2019ThePower}, and more. Giving an account of how social change is fought for, including instances of campaigns for technological abolition or regulation, allows data science ethics to operate with a rich vocabulary of potential action to combat unethical data technology and practices.

A data science student who is equipped with sharp understandings of these areas of social study will be able to ethically reason about technological systems in a holistic and general manner. These are the lines of thinking that are motivating the most influential critiques of data technology, and they should be explicitly taught so data science students leave the university with the skills necessary to engage in the same manner of critique.

\section{Teaching critical data studies through surveillance}

We took these five sub-topics of critical theory and used them to craft a course designed to teach critical data science skills to students. The course analyzed these critical areas through the lens of one sociotechnical system: \emph{surveillance}. 

Surveillance is a growing area of concern in data studies. Zuboff argues that surveillance is the new dominant logic of capital accumulation in the digital economy \cite{Zuboff2019ThePower}. Moreover, major world powers, notably the U.S. and China, are posturing to increase state and economic power by partnering with private technology companies to build ``smart'' borders \cite{DelaHoz2021WhyWall}, cities \cite{Klein2020ScreenDystopia,Andersen2020TheHere}, and militaries (including police forces) \cite{Brewster2021ProjectPentagon,Schmidt2021FinalReport,2018HarnessingProsperity} which operate via surveillance logics. With surveillance systems taking on roles in both public and private life, data science students are likely to grapple with ethical questions surrounding surveillance regardless of their career specialty. By interrogating surveillance technology, we allowed students to build critical skills through the analysis of technology that is in the social milieu, and relevant to their current and future lives.

We additionally note that the choice to focus on surveillance \emph{only} is an intentional choice. Many ethics courses take a broad-view approach, surveying several different types of technologies, perhaps with the intention of teaching students generalizable ethical concerns that cut across these different technologies. One reason we purposefully discussed only surveillance technologies is that these technologies contain all aspects of critical thought we find important. For example, facial recognition technology can be critiqued along the lines of racial outcomes (algorithmic failures to detect dark faces, or its use on marginalized communities by police), the capitalistic forces driving its creation (collecting faces for profit-driven data analysis), the power-brokers who create and use it, its genealogical roots in cameras or CCTV, and reimagined through several theories of change. This is just one type of technology, but it already contains a rich set of critical topics. Moreover, we believe that deep analysis -- seeing that one system is involved in a branching network of concerns, and exploring each of those branches -- is a valuable skill for students. Deep analysis allows students to see complexity, grapple with the interconnectedness of systems, and learn to take time with analysis.

The course was structured to encourage students to build critical skills necessary to analyze potential and actual harms of social technologies. We progressed from deconstructing the \emph{problem} of surveillance technology, to identifying \emph{practices} that enabled that deconstruction, and finally, allowed students to \emph{apply} those practices to a surveillance technology of their own choosing. 

\begin{table*}[t]
\small\sf\centering
\caption{Majors and basic demographics of students who consented through IRB to have their class work be used in this research.}
\begin{tabular}{m{0.1\linewidth}<{\centering} m{0.1\linewidth}<{\centering} m{0.1\linewidth}<{\centering} m{0.1\linewidth}<{\centering} m{0.1\linewidth}<{\centering} m{0.1\linewidth}<{\centering} m{0.1\linewidth}<{\centering}}
\toprule
    & \textbf{Total} & \textbf{Women} & \textbf{Nonbinary} & \textbf{Men} & \textbf{Students of Color} & \textbf{White Students}\\
    \midrule
    STEM Majors & 14 & 7 & 1 & 6 & 11 & 3\\
    \midrule
    Liberal Arts Majors & 7 & 6 & 0 & 1 & 2 & 5\\
\bottomrule
\label{tab:student-demographics}
\end{tabular}
\end{table*}

\subsection{Problem}

The course first deconstructed surveillance by asking what the problem even was in the first place. Students were presented with readings and materials that commented on surveillance through the lenses of capitalism \cite{Zuboff2019ThePower}, racializiation \cite{Browne2015DarkBlackness}, labor relations \cite{Guendelsberger2019OnInsane}, state power \cite{mijente2018WhosBehind,Foucault1975DisciplinePunish}, and border militarization \cite{mijente2018WhosBehind}, as well being presented with a review of surveillance studies in the humanities \cite{Browne2015DarkBlackness}. This type of content gave students tools that they could use to understand the impacts and ethical risks of specific technologies, as well as relevant, contemporary examples of the usage of surveillance technology and resulting harms.

This section also historicized surveillance technology through Browne's \emph{Dark Matters} \cite{Browne2015DarkBlackness}. Browne discusses surveillance in the early United States designed to control enslaved people, depicting objects like slave ships and wanted ads as surveillance technologies. Her extended discussion of New York City's ``lantern laws'' was memorable for many students -- the laws being that Black, Brown and Indigenous people were required to illuminate themselves with lantern light at night as a means to police them. These examples challenged students' conception of technology, demonstrating that even a lantern can be a surveillance technology, and also pushed them to see that social systems contribute to unethical technology as much as the devices themselves.

To conclude the section, we examined acts of resistance to surveillance technology and touched on theories of change. Students engaged with material depicting the resistance of enslaved people to surveillance technology used for slavery \cite{Browne2015DarkBlackness}, worker subversion of the surveillance of scientific management \cite{Guendelsberger2019OnInsane}, the Alphabet Workers' Union's resistance to unethical projects \cite{logic2021NowIKnow}, and student activism against Big Tech involvement with Immigration and Customs Enforcement's (ICE) surveillance aimed at deportation \cite{mijente2018WhosBehind}.

We also gathered a panel of local organizers to discuss their work and field questions from students, including an employee of the American Civil Liberties Union (ACLU) who fought facial recognition proliferation in Massachusetts, a Somerville City Councilor at Large, and a Harvard graduate student active in student organizing. 

\subsection{Practice}

The course then pulled the lens back and examined data ethics through abstract lenses of power, analysis, and design. This section was intended to allow students to explicitly put words to the practice that was just used during the \emph{problem} section.

Students interfaced with texts from \emph{Data Feminism} \cite{DIgnazio2020DataFeminism} and \emph{Design Justice} \cite{Costanza-Chock2020DesignNeed}, as well as a talk from Dr. Ruha Benjamin from the \emph{International Conference on Learning Representations} \cite{Benjamin20202020Society}. Through these materials, they learned about power and its interaction with systems of oppression -- processes that lead data-driven technologies to recreate those same systems of oppression. They learned about the way that data and data sets are subject to social forces, making them a representation of the biases that their creators hold.

The class was exposed to several lines of critical questioning: asking who holds power in society and how that affects technology design, identifying several systems of oppression and asking how they manifest in a technology, inquiring as to who benefits and who is harmed by a technology, asking who is at the helm of design, questioning who is resisting any harms, and wondering what a technology would look like reimagined. 



\subsection{Application}

Finally, the course culminated in students putting their newfound knowledge and critical skills to the test. The class analyzed 15 different surveillance measures in groups and presented their work to a public exhibition at Tufts University, with some also submitting written reports.

Students were guided by a brief prompt to scaffold their projects. The questions presented to students included: (1) What is the surveillance measure and what organization(s)/company(ies) created and implemented it? (2) What is the ‘problem’ this surveillance measure set out to solve? (3) What are the historical \& present contexts in which this surveillance measure came to be? (4) What are the benefits and harms of this surveillance measure? For whom? By whom? (5) What does resistance look like for this surveillance measure? (6) What would this surveillance measure (or context) look like reimagined?

Through their reflections and final projects, students in the course continually demonstrated that they were learning a method of ethical thinking in a data-driven world that extends beyond presentation of case studies or consideration of traditional ethical frameworks. They showed an understanding of social context and flexibility in analyzing contemporary surveillance technologies that allowed them to leverage smart and nuanced critiques spanning several dimensions of analysis.

\section{Evidence of critical and ethical learning}

\subsection{Research questions and methods}

We were primarily motivated by two major research questions:

\begin{itemize}
    \item RQ1: Is there evidence of student learning along the five sub-topics of critical theory (oppression, social logics, power, history, and change)?\\
    \item RQ2: Is there evidence of student ability to critically analyze technologies with minimal scaffolding (essentially generating case studies)?
\end{itemize}

We used student assignment responses as data to investigate these two research questions. Throughout the course, students wrote weekly reflections, guided by prompts, including a final reflection on their learning throughout the course at the end of the semester. Student responses were collected from those who consented, through the Tufts IRB, to allow their class writing to be analyzed and used in research. In total, 25 of 38 students consented to have their assignment responses used in research, and we focus on 21 of the 25 consenting students' final reflection responses. A breakdown of students by demographics and major is displayed in Table \ref{tab:student-demographics}. The 6 students we omit did not hand in their final reflection in time for the analysis performed.

To draw out themes that appeared in student writing, we reviewed all student reflections and used an iterative inductive coding process \cite{OReilly2012EthnographicEdition} -- reading line-by-line and using open coding (themes discovered in text, added to a code book, and then searched for in students responses in subsequent coding iteration, until satisfaction was reached). Both authors reached agreement on each theme and their presence in student responses. Author dialogue was important to address biases in individual readings of student responses \cite{Creswell2000DeterminingInquiry}.

The major themes of student learning we arrived at through our coding process were: (1) relation of technology and society, (2) social values and logics, (3) ethical skills, (4) student motivation to act ethically, and (5) interdisciplinary or ``different'' class environment. The number of student final reflections which contained writing addressing these themes is displayed in Table \ref{tab:inductive-themes}. In \cite{Ozkan2024CriticalEducation} we examined student responses along these five themes, but in this work, we instead focus on 2, 3, and 4, and put them in context of the five sub-topics of critical theory outlined in the section above.

\begin{table*}[t]
\small\sf\centering
\caption{Results of inductive coding of student final reflection assignment responses, disaggregated by student major.}
\begin{tabular}{m{0.1\textwidth}<{\centering} m{0.1\textwidth}<{\centering} m{0.1\textwidth}<{\centering} m{0.1\textwidth}<{\centering} m{0.1\textwidth}<{\centering} m{0.1\textwidth}<{\centering} m{0.2\textwidth}<{\centering}}
\toprule
    & \textbf{Total} & \textbf{Relation of Technology \& Society} & \textbf{Social Values \& Logics} & \textbf{Ethical Skills} & \textbf{Student Motivation to Act Ethically} & \textbf{Interdisciplinary or ``Different'' Class Environment}\\
    \midrule
    STEM Majors & 14 & 12 (85\%) & 9 (63\%) & 10 (71\%) & 9 (63\%) & 7 (50\%) \\
    \midrule
    Liberal Arts Majors & 7 & 2 (29\%) & 4 (57\%) & 3 (43\%) & 7 (100\%) & 4 (57\%)\\
\bottomrule
\label{tab:inductive-themes}
\end{tabular}
\end{table*}

\begin{table*}[t]
\small\sf\centering
\caption{Results of re-coding of student final reflection assignment responses falling in ``Social Values \& Logics,'' ``Ethical Skills,'' and ``Student Motivation to Act Ethically'' (20 of 21 students) according to the five sub-topics of critical theory motivating the article, disaggregated by student major.}
\begin{tabular}{m{0.1\textwidth}<{\centering} m{0.1\textwidth}<{\centering} m{0.1\textwidth}<{\centering} m{0.1\textwidth}<{\centering} m{0.1\textwidth}<{\centering} m{0.1\textwidth}<{\centering} m{0.1\textwidth}<{\centering}}
\toprule
    & \textbf{Total} & \textbf{Oppression} & \textbf{Social Logics} & \textbf{Power} & \textbf{History} & \textbf{Change}\\
    \midrule
    STEM Majors & 14 & 8 (57\%) & 6 (43\%) & 6 (43\%) & 3 (21\%) & 6 (43\%) \\
    \midrule
    Liberal Arts Majors & 6 & 3 (50\%) & 2 (33\%) &  1 (17\%) & 0 (0\%) & 4 (66\%)\\
\bottomrule
\label{tab:critical-topic-coding}
\end{tabular}
\end{table*}

In this work, the first author performed a subsequent qualitative coding of student responses within themes 2, 3, and 4 (20 of 21 students) along the five sub-topics of critical theory. The results of this coding are listed in Table \ref{tab:critical-topic-coding}. Below, we use the data from student reflections, as well as examples from student final projects, to explore our research questions and gauge student learning.


\subsection{Critical theory skills}

In our coding of student final reflections, we found many examples of students discussing social values and logics, ethical skills, and the motivation to act, that correspond to our learning goals across the five critical theory areas. From the data in Table \ref{tab:critical-topic-coding}, we see that the most discussed areas are, in order: oppression, change, social logics, power, and history.

Within students' discussion of oppression, the most common topic mentioned was racializing surveillance. One humanities student shared her experience of post-9/11 surveillance in New York City as an Arab-American. She wrote that, ``I came into into this course with a very clear understanding of the real world implications that racialized surveillance can have on the health and livelihoods of communities of color. What I didn't have, however, was the vocabulary, background readings, and technical ability to call out this surveillance, label and define it, and pose ways to evade and resist it.'' Another student, a design major, shared that ``Browne's readings helped me think more about the intersections of race and technology, and how they personally affect me and my communities as a Latinx woman.'' We find it notable that these students' lived experiences were given specific vocabulary in the classroom, which they found empowering as a means to connect technology with their lives and make sense of their worlds.

In students' writing about social logics, the most prevalent topics were capitalism and surveillance capitalism. Several students expressed their shock when reflecting on how they learned the extent to which their social media or internet usage was tracked and used to serve them ads or predict their behavior. One computer science student shared: ``I have become more conscious about my digital footprint and the personal data that I offer online, being more careful to avoid activities that can lead to me bring tracked.'' A humanities student noted that ``this class has totally shifted the way I understand technology in my world, the way I use the internet, and understanding how my data contributes to surveillance capitalism.''

When students discussed capitalism, they largely expressed displeasure with being controlled by large companies, and the use of technology not for social good but for profit. One design student critiqued the system, arguing that ``the main purpose of technology is not the improvement of the lives of humanity, but to make money for the few in the top percent.'' A computer science major shared that her conception of surveillance expanded to include ``the capitalistic structures and frameworks that empower those with money and capital while pitting workers against each other and dividing them.''

Theories of change were also discussed by many students. They uplifted several types of change work, including resistance, collective action, community organizing, and social justice organizations. Several students also specifically mentioned the panel of local change-makers that we featured early in the semester. A computer science student mentioned that, ``the panel of organizers... stuck with me because it made connections from the larger themes of the course down to a local level.'' Another computer science student wrote that, ``I changed my view of resistance after I saw how [the panelists] and so many others find the injustice in our neighborhood. How they use their individual power to create a larger impact and some even change the law toward a better direction.''

When students wrote about power, they predominantly mentioned it in the abstract, as they learned to see more clearly that some wield control over others. Many students mentioned the incentives leading organizations and companies to surveil, for profit or power, and articulated their sense of injustice at some using such technologies for these purposes.

The few students who discussed history mostly mentioned Browne's discussion of lantern laws, specifically as they connected to modern issues of surveillance against marginalized groups. The design student who felt her lived experience resonate with discussions of racializing surveillance also wrote that ``Reading about the ways early technologies were used to surveil enslaved Black people, with items as simple as lamplights, to the ways they are used in modern day Palestine, was enlightening.''

In total, we find strong evidence towards answering our first research question in the positive: student did internalize methods of thinking and critique along the five sub-topics of critical theory. Some aspects of critique stuck more with students than others, which can be addressed in future instruction by adding other examples that lend themselves to these less-discussed topics.

\subsection{Generative ethical skills}


To gauge our second research question, which focuses on students' abilities to perform analysis in a manner that is \emph{generative} of ethical cases, we analyze student final projects. Through these projects, students demonstrated their capacity to examine their own world and find ethical issues. Much like the way that now-famous case studies of unethical data science arose, students used their critical skills to critique surveillance systems, reimagine alternatives, and present their findings to their peers.

As it is difficult to share all student projects within the space confines of this article, we uplift and describe three examples of excellent student projects. Each of them picked very different systems to analyze, but all demonstrated strong skills in critiquing them. 




One student group focused on the Amazon Ring doorbell, which boasts a camera enabling consumers to see those ringing the doorbell and potentially catch any attempts at package robbery. The group was composed of one computer science, one environmental engineering, and one women's gender and sexuality studies student. The interdisciplinary group incorporated surveillance theories from Foucault's writing on Bentham's panopticon to argue that Ring oversteps its stated goal of safety -- rather, embodying an all-seeing system that is too powerful. They critique that power by reporting Amazon's widepsread partnerships with police, who obtain user data from Amazon through warrants, and wield what they see as ``government-like power'' in an unaccountable, undemocratic way. They go further to critique the differential impacts of facial recognition technology that could be used in conjunction with the Ring's video feeds, which have been shown to misidentify people of color, especially Black people, and has led to false accusations of crime by police.
From this research, the group generated several vignettes to be used as ethical case studies, evidence that learning the abstract theories affecting technological systems allowed them to identify specific potential harms in a more flexible manner than starting with case studies alone.

Another group, comprised of two computer science students and one science, technology and society student, analyzed the Saudi-sponsored smart city, NEOM, and its usage of surveillance technology. Their final report included critiques of the city's stated goals: boasting improvements to life via data collected inside the city (including personalized entertainment, data-driven healthcare, and digital financial systems), within a city powered by autonomous vehicles and carbon-neutral electricity. Yet the group was able to cut through the rhetoric, asking whose life is actually enhanced by the city. They reported the history of the land NEOM is being built on, which has been inhabited by indigenous people of the Huwaitat tribe, and is now displacing them to build the city.
They additionally critique the beneficence of NEOM's healthcare claims, pointing out that the city's pursuit of care based on genomics and genetics could easily be used to collect surplus data and monetize sensitive information without consent. This group demonstrated a critical ability to use the concepts of surveillance capitalism and state use of surveillance to see through a shiny, buzzword-rich technology which claims to benefit humanity, but can be shown to be fraught.

Finally, a group of three computer science students examined the technology ShotSpotter, which claims to listen for gunshots around cities and quickly alert police to potential crimes. The group points out issues with the technology stemming from the racist culture it is deployed in, reporting that ShotSpotter is mostly used in minority communities, and leads to over-policing. They reference its over 70\% false positive rate to argue that it does not reduce crime. They additionally note that police departments have adopted it en masse at the cost of \$100,000 per square mile, per year. Part of their project consisted of a simulation where participants could apportion a municipal budget to attempt to reduce crime in a digital city. They could choose several programs to purchase, including a youth jobs program, emergency aid, cleaning up vacant lots, or licensing ShotSpotter. Their project demonstrated their understanding that the benefits technologies purport to bring to the social world can also be brought about by social programs -- ones which may have far fewer unintended side-effects, are more accountable, and are more effective.

\section{Discussion and future work}

Through designing a course that critically examined surveillance technology through several social theoretical dimensions -- systems of oppression, social logics, considerations of power, historical analysis, and theories of change -- we were able to make progress towards addressing a crucial gap in data science ethics instruction. Typical data science ethics, and engineering ethics more largely, are taught in terms of philosophical theories of ethics, or through individualized ethical dilemma case studies -- methods whose conclusions are subject to models of the world derived from social understandings. This course aimed at teaching those social understandings so that data science students could complement typical ethical study with understandings of what makes situations ethical or unethical in the first place.

We found strong evidence of students' ability to critique technology at an abstract and concrete level in their reflection assignments and final projects. Our analysis of student work throughout the course showed that students were understanding and utilizing five aspects of critical theory that we found important to critical data science thinking: oppression, social logics, power, history, and theories of change. Moreover, we found through student final projects that with the help of gentle guidance and scaffolding, students successfully critiqued systems relevant to their own lives, essentially generating new case studies.

This model of ethics instruction can be utilized in other curricula to build critical skills for students. We believe that this type of ethics instruction complements trends using moral philosophical theories for decision-making, as well as the move to prioritize fairness, accountability, and transparency in algorithmic design. The degree to which our institutional context maps onto others may be limited, and should be a topic of future study. Additionally, more research can be conducted to gauge the degree to which these critical thinking skills, particularly in the realm of theories of change, can be applied by former students who are now practitioners in industry settings.

\bibliographystyle{plainnat}

\end{document}